\documentclass[
  journal=pasa,
  manuscript=research-paper,
  year=202X,
  volume=XX,
]{cup-journal}

\usepackage{amsmath}
\usepackage[nopatch]{microtype}
\usepackage{booktabs}
\usepackage{multirow}
\usepackage{gensymb}

\usepackage{url}
\usepackage{hyperref} 

\def\BI{\hbox{B2\,{\sc i}-{\sc iii}}}
\def\BII{\hbox{B2\,{\sc ii}}}
\def\BIII{\hbox{B2\,{\sc iii}-{\sc iv}}}
\def\HII{\hbox{H{\sc ii}}}

\newcommand{\farcm}{\mbox{\ensuremath{.\mkern-4mu^\prime}}}
\newcommand{\farcs}{\mbox{\ensuremath{.\!\!^{\prime\prime}}}}
\newcommand{\fdg}{\mbox{\ensuremath{.\!\!^\circ}}}

\usepackage[nolist]{acronym}
\begin{acronym}[AWGN]
\acro{ASKAP}{Australian Square Kilometre Array Pathfinder}
\acro{CARTA}{Cube Analysis and Rendering Tool for Astronomy}
\acro{EMU}{Evolutionary Map of the Universe}
\acro{RN}{Reflection Nebula}
\acrodefplural{RN}[RNe]{Reflection Nebulae}
\acro{RNe}{Reflection Nebulae}
\acro{DR3}{Data Release 3}
\acro{PM}{proper motion}
\acro{CO}{carbon monoxide}
\acro{CASDA}{CSIRO ASKAP Science Data Archive}
\acro{RMS}{Root Mean Squared}
\acro{2MASS}{Two Micron All Sky Survey}
\acro{DSS}{Digitised Sky Survey}
\acro{MIRIAD}[\textsc{Miriad}]{Multichannel image reconstruction, image analysis and display}
\acro{ZAMS}{zero-age-main-sequence}
\acro{WISE}{Wide-Field Infrared Survey Explorer}
\acro{NVSS}{NRAO VLA Sky Survey}
\end{acronym}

\title{ASKAP EMU Radio Detection of the Reflection Nebula VdB-80 in the Monoceros Crossbones Filamentary Structure}

\author{A. C. Bradley}
\affiliation{Western Sydney University, Locked Bag 1797, Penrith South DC, NSW 2751, Australia}
\email[A. C. Bradley]{20295208@student.westernsydney.edu.au}

\author{Z. J. Smeaton}
\affiliation{Western Sydney University, Locked Bag 1797, Penrith South DC, NSW 2751, Australia}

\author{N. F. H. Tothill}
\affiliation{Western Sydney University, Locked Bag 1797, Penrith South DC, NSW 2751, Australia}

\author{M. D. Filipovi\'c}
\affiliation{Western Sydney University, Locked Bag 1797, Penrith South DC, NSW 2751, Australia}

\author{W. Becker}
\affiliation{Max-Planck Institut f\"ur extraterrestrische Physik, Gie\ss enbachstra\ss e 1, 85748 Garching, Germany}
\alsoaffiliation{Max-Planck-Institut f\"ur Radioastronomie, Auf dem H\"ugel 69, 53121 Bonn, Germany}

\author{A. M. Hopkins}
\affiliation{School of Mathematical and Physical Sciences, 12 Wally's Walk, Macquarie University, NSW 2109, Australia}

\author{B. S. Koribalski}
\affiliation{Australia Telescope National Facility, CSIRO, Space and Astronomy, PO Box 76, Epping, NSW 1710, Australia}
\alsoaffiliation{Western Sydney University, Locked Bag 1797, Penrith South DC, NSW 2751, Australia}

\author{S. Lazarevi\'c}
\affiliation{Western Sydney University, Locked Bag 1797, Penrith South DC, NSW 2751, Australia}
\alsoaffiliation{Australia Telescope National Facility, CSIRO, Space and Astronomy, PO Box 76, Epping, NSW 1710, Australia}
\alsoaffiliation{Astronomical Observatory, Volgina 7, 11060 Belgrade, Serbia}

\author{D. Leahy}
\affiliation{Dept. of Physics and Astronomy, University of Calgary, Calgary, AB, T2N 1N4, Canada}

\author{G. Rowell}
\affiliation{School of Physics, Chemistry and Earth Sciences, The University of Adelaide, Adelaide 5005, Australia}

\author{V. Velovi\'c}
\affiliation{Western Sydney University, Locked Bag 1797, Penrith South DC, NSW 2751, Australia}

\author{D. Uro\v{s}evi\'c}
\affiliation{Department of Astronomy, Faculty of Mathematics, University of Belgrade, Studentski trg 16, 11000 Belgrade, Serbia}


\begin{document}

\begin{abstract}
 We present a new radio detection from the \ac{ASKAP} \ac{EMU} survey associated with the \ac{RN} VdB-80. The radio detection is determined to be a previously 
 unidentified \HII\ region, now named \textit{Lagotis}. 
 The \ac{RN} is located 
 towards Monoceros,
centred in the molecular cloud feature known as the `Crossbones'. 
 The 944\~MHz \ac{EMU} image shows a roughly semicircular \HII\ region with an integrated flux density of 30.2$\pm$0.3\,mJy. The \HII\ region is also seen at 1.4~GHz by \ac{NVSS}, yielding an estimated  spectral index of 0.65$\pm$0.51, consistent with thermal radio emission. \textit{Gaia} \ac{DR3} and \ac{2MASS} data give a distance to the stars associated with the \HII\ region of $\sim$960~pc. This implies a size of 0.76$\times$0.68($\pm$0.09)\,pc for the \HII\ region.
 We derive an \HII\ region electron density of 
 the bright radio feature to be 
 26\,cm$^{-3}$, requiring a Lyman-alpha photon flux of 
 $10^{45.6}$
 \,s$^{-1}$, which is consistent with the expected Lyman flux of HD\,46060, the \BII\ type star which is the likely ionising star of the region. The derived distance to this region implies that the Crossbones feature is a superposition of two filamentary clouds, with Lagotis embedded in the far cloud.
\end{abstract}

\acresetall 

\section{Introduction}
\label{sec:Introduction}

\acp{RN} are diffuse clouds of gas and dust typically associated with star-forming regions. These \acp{RN} are illuminated by young stars and may be accompanied by emission nebulae \citep{2000PASP..112.1426M, 2024MNRAS.529.1680E}. These nebulae are prominent in visible wavelengths, but the visible emission of a \ac{RN} may be obstructed by dark molecular gas and dust, so the full extent is often better seen in  near-infrared \citep{1981ApJ...246..416W} and far-infrared \citep{1984ApJ...277..623S} wavelengths. 

VdB-80 (also known as [RK68]-59), is a known Galactic \ac{RN} in the constellation Monoceros, below the Galactic plane at $l$\,=\,219$\fdg$26, $b$\,=\,--8$\fdg$93 \citep[RA(J2000) $\sim 06^{\mathrm h}31^{\mathrm m}$, DEC(J2000) $\sim -9^\circ39^\prime$,][] {1966AJ.....71..990V, 1968TrAlm..11....3R, 2003A&A...399..141M}.

\indent\HII\ regions are typically associated with sites of early star formation or young stellar clusters embedded in molecular clouds and are comprised of ionised hydrogen \citep{1997pism.book.....D}. \citet{2001A&A...377..845A} did not see any signs of an \HII\ region towards VdB-80 in their optical spectra.
Nonetheless, we identify radio emission detected by ASKAP as an \HII\ region that we name `Lagotis'\footnote{Named after the Australian Greater Bilby (see section \ref{subsec:HD46060}).}, 
associated with a stellar cluster in the `Crossbones' molecular cloud, and it is proposed that the star HD~46060 is the centre of both \ac{RN} and \HII\ region.

The Crossbones is a filamentary cloud structure located within the Mon R2 complex, first characterised by \citet{1986ApJ...303..375M}, potentially related to the Orion-Eridanus superbubble \citep{2009ApJ...694.1423L}. This `X' shaped structure is an active star-forming region, with VdB-80 bordering the cloud near to its centre. The structure is most clear in maps of the (J = 1--0) rotational transition of $^{12}$CO \citep{1987ApJ...322..706D, 2024A&A...688A..54G}.

The \ac{ASKAP} \citep{2021PASA...38....9H} \ac{EMU} \citep[Hopkins et al., PASA, submitted.]{Norris2011,Norris2021} survey has provided new radio-continuum observations of this region with improved sensitivity compared to any previous radio-continuum surveys. Higher sensitivity has allowed for the first reliable detection of radio-continuum emission toward VdB-80. This new radio detection allows us to estimate properties of the \ac{RN} that confirm the presence of a \HII\ region, as well as confirming HD~46060's role as the star responsible for creating both features.

This radio detection demonstrates the ability of the newest generation of radio surveys, such as \ac{EMU}, to detect new low surface-brightness emission. This has been demonstrated in several new discoveries, such as supernova remnants G181.1--9.5~\citep{2017A&A...597A.116K}, J0624--6948~\citep{2022MNRAS.512..265F}, G288.8--6.3~\citep[Ancora;][]{2023AJ....166..149F,2024A&A...684A.150B}, G121.1--1.9 \citep{2023MNRAS.521.5536K}, G308.7+1.4 \citep[Raspberry;][]{2024RNAAS...8..107L}, G312.6+ 2.8 \citep[Unicycle;][]{2024RNAAS...8..158S}, and a pulsar wind nebula \citep[Potoroo;][]{2024PASA...41...32L}.

The paper is structured as follows: Section \ref{sec:Data} outlines the data used, including radio, optical, and infrared data, both new and archival. Section \ref{sec:Results and Discussion} provides analysis and interpretation of Lagotis, this includes distance, size, emission properties and the context of the \ac{RN} within the Crossbones filaments. Section \ref{sec:Conclusion} provides a brief summary of our results and interpretation of Lagotis.

\section{Data} 
\label{sec:Data}
\subsection{Radio observations}
\label{subsec:Radio}
\subsubsection{ASKAP} 
\label{subsubsec:ASKAP}
\indent

The radio emission associated with VdB-80 was discovered in the \ac{ASKAP} \ac{EMU} (AS201) survey. 
Lagotis appears in two separate \ac{EMU} observations: Scheduling Block SB59692 observed tile EMU$\_0626-09$A on 2024 March 02, and SB61077 observed tile EMU$\_0626-09$B on 2024 April 13. Both observations were taken at 943.5\,MHz with a bandwidth of 288\,MHz. The data were reduced using the standard \ac{ASKAP} pipeline, ASKAPSoft, using multi-frequency synthesis imaging, multi-scale cleaning, self-calibration and convolution to a common beam size \citep{2019ascl.soft12003G}. 

\begin{figure*}[ht]
\centerline{\includegraphics[trim=0 20 0 15 width=1\linewidth, keepaspectratio]{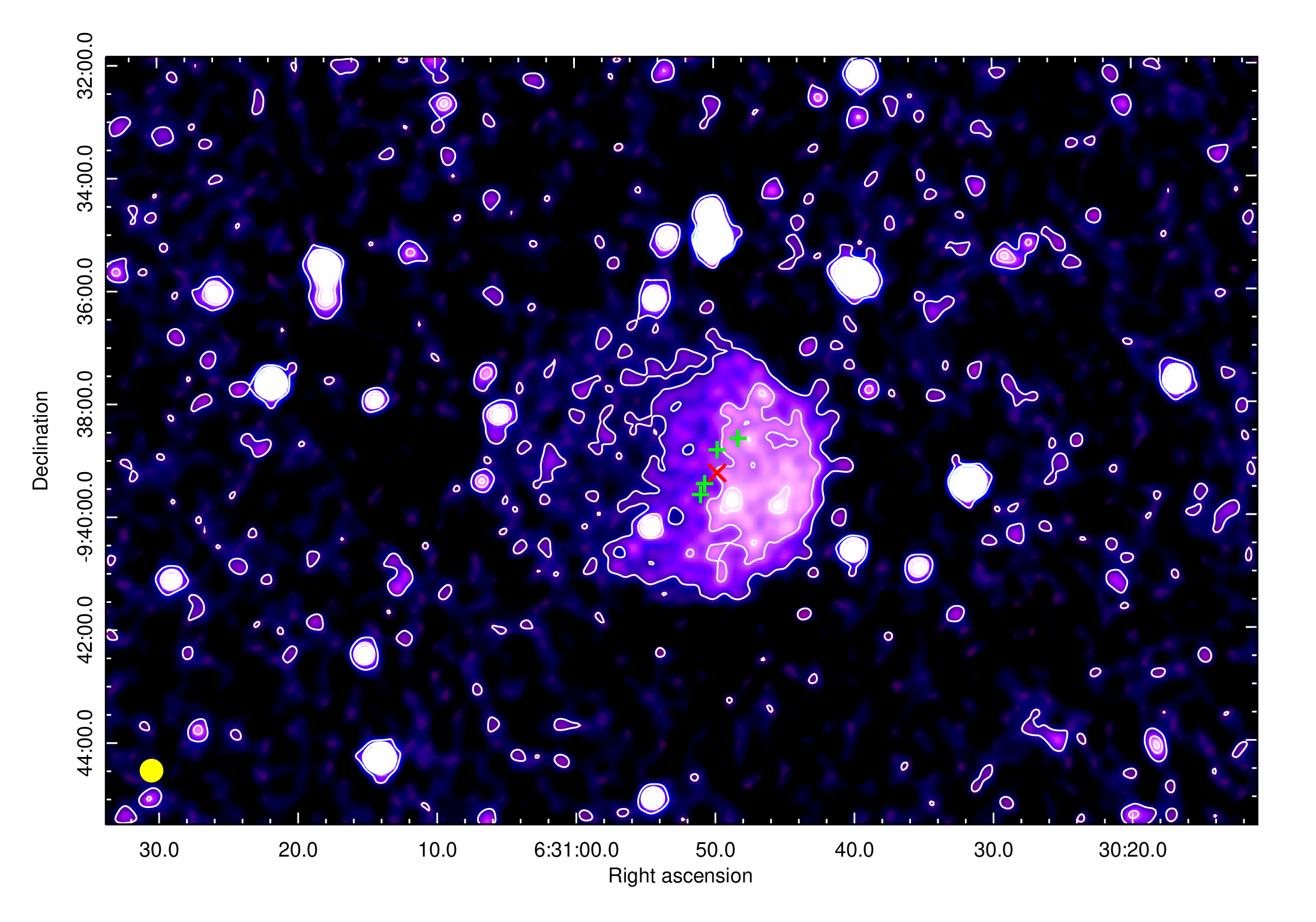}}
\caption{EMU radio-continuum image of Lagotis and VdB-80 at 944\,MHz. Local RMS noise is 20$\,\mu$Jy beam$^{-1}$, and contours are at levels of 3, 10 and 15$\sigma$. The image resolution is 15$\times$15\farcs, represented with the yellow circle in the bottom left corner. The red `X' denotes the star HD~46060, and the green crosses denote the other stars in the cluster (see Table \ref{table:GaiaData}).}
\label{fig:Lagotis1}
\end{figure*}

We merged the two images using \ac{MIRIAD}~\citep{Sault1995} task \textsc{imcomb}. This merges the images together while applying weighting in the overlapped region to minimise rms noise. SB59692 had significantly more interference from a nearby bright star, and so the weighting ratio of 1:1.5, for SB59692:SB61077, was used to produce the best image. The resulting Stokes~$I$ image (Figure~\ref{fig:Lagotis1}) has a restored beam size of 15$\times$15\farcs and a local rms noise of $\sim$20\,$\mu$Jy\,beam$^{-1}$. There is no corresponding Stokes~$V$ emission.

\subsubsection{NVSS}
\label{subsubsec:NVSS}
\indent

The \ac{NVSS} covers the entire sky north of $-40^\circ$ \citep{1998AJ....115.1693C}, at 1.4~GHz in the radio continuum, with bandwidth 50~MHz. The Lagotis \HII\ region is also present in this survey, which is used in section \ref{subsec:RCE} to obtain characteristics of the \ac{RN} in radio wavelengths. The \ac{NVSS} image used has a resolution of 45$^{\prime \prime}\times$45$^{\prime \prime}$ and a measured local rms noise of $\sim$40$\mu$Jy beam$^{-1}$.

\subsubsection{CO data}
\label{subsubsec:CO}
\indent

The Crossbones cloud was first observed by \citet{1986ApJ...303..375M} as part of the Columbia CO survey of the galaxy \citep{1987ApJ...322..706D}. Newer $^{12}$CO ($J=1\rightarrow 0$) maps of the Galaxy have been derived from the broadband millimetre-wave maps   obtained by the \emph{Planck}
satellite \citep{2024A&A...688A..54G}.


\subsection{Optical observations}
\label{subsec:Optical}

\citet{1966AJ.....71..990V} identified HD\,46060 as the illuminating star of the \ac{RN}, noting that it is ``in a small compact clustering which includes BD\,$-9^\circ\, 1497$.'' \citet{2015NewA...36...70O} used the UCAC4 astrometric catalogue to identify 8 cluster members of the 23 proposed by \citet{2009MNRAS.397.1915B}.

\subsubsection{Gaia} 
\label{subsubsec:Gaia}
\indent

\textit{Gaia} data for 5 stars that appear to lie within the radio-continuum emission were taken from the \textit{Gaia} \ac{DR3} \citep{2016A&A...595A...1G, 2023A&A...674A...1G} catalogue. These values, as well as derived distances, are reported in Table \ref{table:GaiaData}. Three of these stars are also listed by \citet{2015NewA...36...70O}; the proper motions are fairly similar, except for UCAC4 402-013691 --- \citeauthor{2015NewA...36...70O} do not consider it to be a cluster member, but the \textit{Gaia} \ac{PM} measurement are much closer to the cluster average.

\begin{table*}
\caption{\textit{Gaia} properties of stars within VdB-80. Columns [2] and [3] are FK5 (J2000) right ascension and declination positions. Column [4] is parallax and its associated error in milliarcseconds. Column [5] is distance and its associated error, as calculated from column [4], in parsecs. Column [6] is the right ascension proper motion, and column [7] is the declination proper motion in milliarcseconds per year.}
\vskip.25cm
\centerline{\begin{tabular}{lcccccc}
 \hline
 [1] & [2] & [3] & [4] & [5] & [6] & [7] \\
 Star Name & RA (J2000) & DEC (J2000) & $p\pm\Delta p$ & $d\pm\Delta d$ & $\mu$ (RA) & $\mu$ (DEC) 
 \\
  & (h:m:s) & (d:m:s) & (mas) & (pc) & \multicolumn{2}{c}{(mas yr$^{-1}$)} 
  \\
 \hline\hline
 HD 46060; NSV\,2998; UCAC4\,402-013688 & 06:30:49.8 & --09:39:14.8 & 1.0711$\pm$0.0220 & 933$\pm$19 & --3.526 & 0.154 
 \\
 BD-09 1497 & 06:30:48.3 & --09:38:38.0 & 1.0532$\pm$0.0191 & 949$\pm$18 & --3.428 & 0.226
 \\
 UCAC4 402-013691 & 06:30:50.7 & --09:39:26.1 & 1.0261$\pm$0.0184 & 975$\pm$18 & --3.520 & 0.730 
 \\
 UCAC4 402-013693 & 06:30:51.0 & --09:39:37.7 & 1.0430$\pm$0.0238 & 959$\pm$22 & --3.190 & 0.270 
 \\
 Gaia DR3 3002950320576187904 & 06:30:49.8 & --09:38:50.1 & 0.9483$\pm$0.0645 & 1055$\pm$72 & --3.210 & 0.199 
\\
\hline
\end{tabular}}
\label{table:GaiaData}
\end{table*}

\subsection{Infrared observations}
\label{subsec:Infrared}

\subsubsection{2MASS}
\label{subsubsec:2MASS}
\indent

Data were collected from \ac{2MASS} \citep{2006AJ....131.1163S} and used in Section \ref{subsec:Distance and Size} to determine accurate distance estimates to the VdB-80, as well as confirming the distance association to the Lagotis \HII\ region. The \ac{2MASS} All Sky Catalogue of point sources provides the magnitudes of the \ac{RN} associated stars in the J, H and K$_\mathrm{s}$ (1.25$\mu$m, 1.65$\mu$m and 2.17$\mu$m) infrared bands. To verify the association of the \ac{RN} and the \HII\ region, these magnitudes, as well as a \ac{2MASS} J--H/H--K$_s$ extinction map from \citet{2007MNRAS.378.1447F}, were used to determine stellar position and dust extinction values. 

\subsubsection{WISE}
\label{subsubsec:WISE}
\indent

\begin{figure*}[ht]
\centerline{\includegraphics[width=1\linewidth, keepaspectratio]{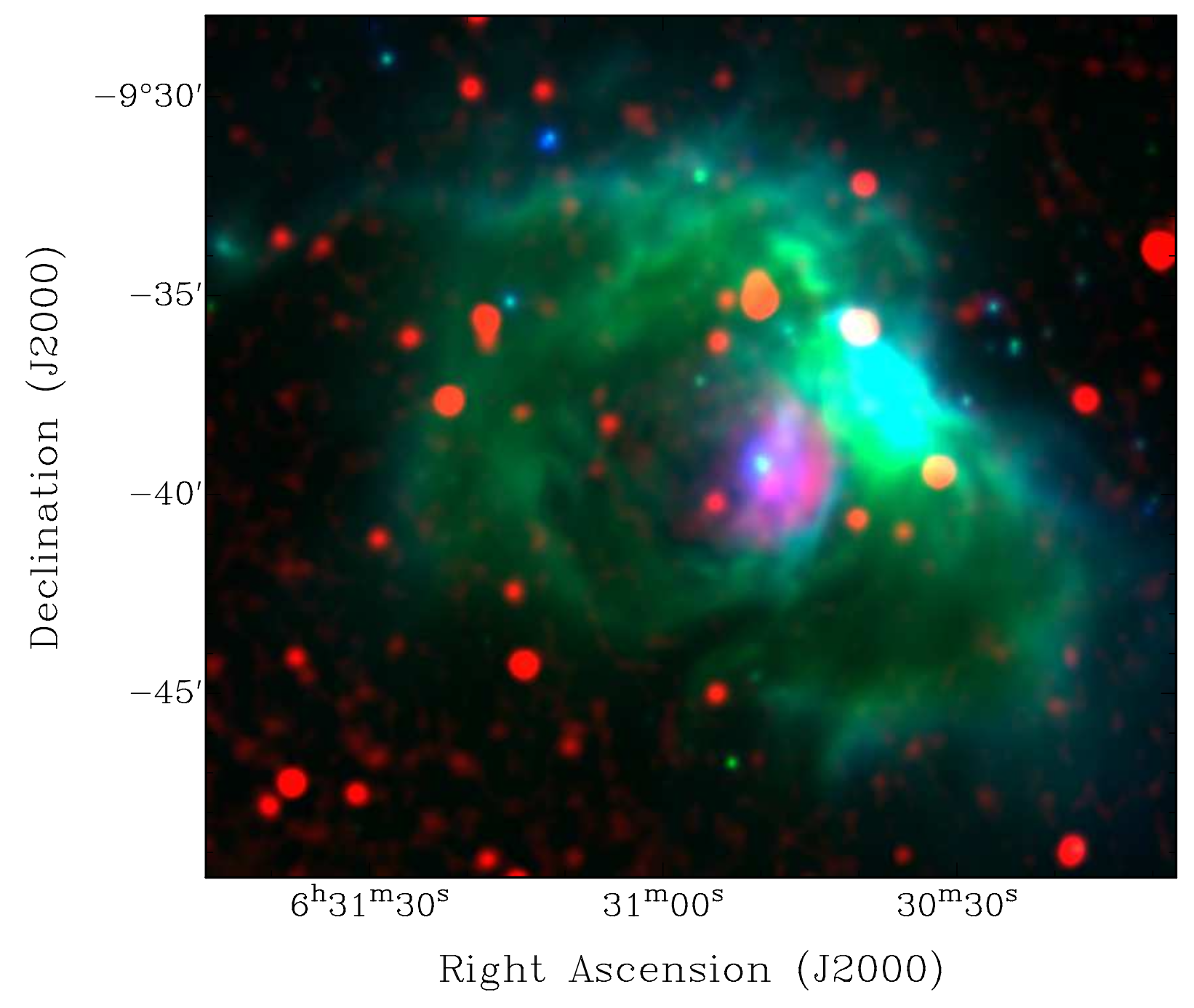}}
\caption{An RGB image tracing the near-infrared emission of VdB-80, where red is the \ac{EMU} tile SB61077 (smoothed to a 25" resolution) at 943.5~GHz, green is the WISE W3 band (12$\mu$m) and blue is the WISE W4 band (22$\mu$m).}
\label{fig:WISE34}
\end{figure*}

The \ac{WISE} is an all-sky survey in near- and mid-infrared \citep{2010AJ....140.1868W}. Observations were taken in four bands, W1, W2, W3, and W4 with wavelengths 3.4$\mu$m, 4.6$\mu$m, 12$\mu$m and 22$\mu$m. The W3 and W4 bands are used with \ac{EMU} data (Figure~\ref{fig:WISE34}) to trace the mid-infrared emitting gas and dust that shows the full extent of the \ac{RN} \citep{book2} at wavelengths other than optical.

\subsubsection{AKARI}
\label{subsubsec:Akari}
\indent

\begin{figure*}[ht!]
\centerline{\includegraphics[width=1.0\linewidth, keepaspectratio]{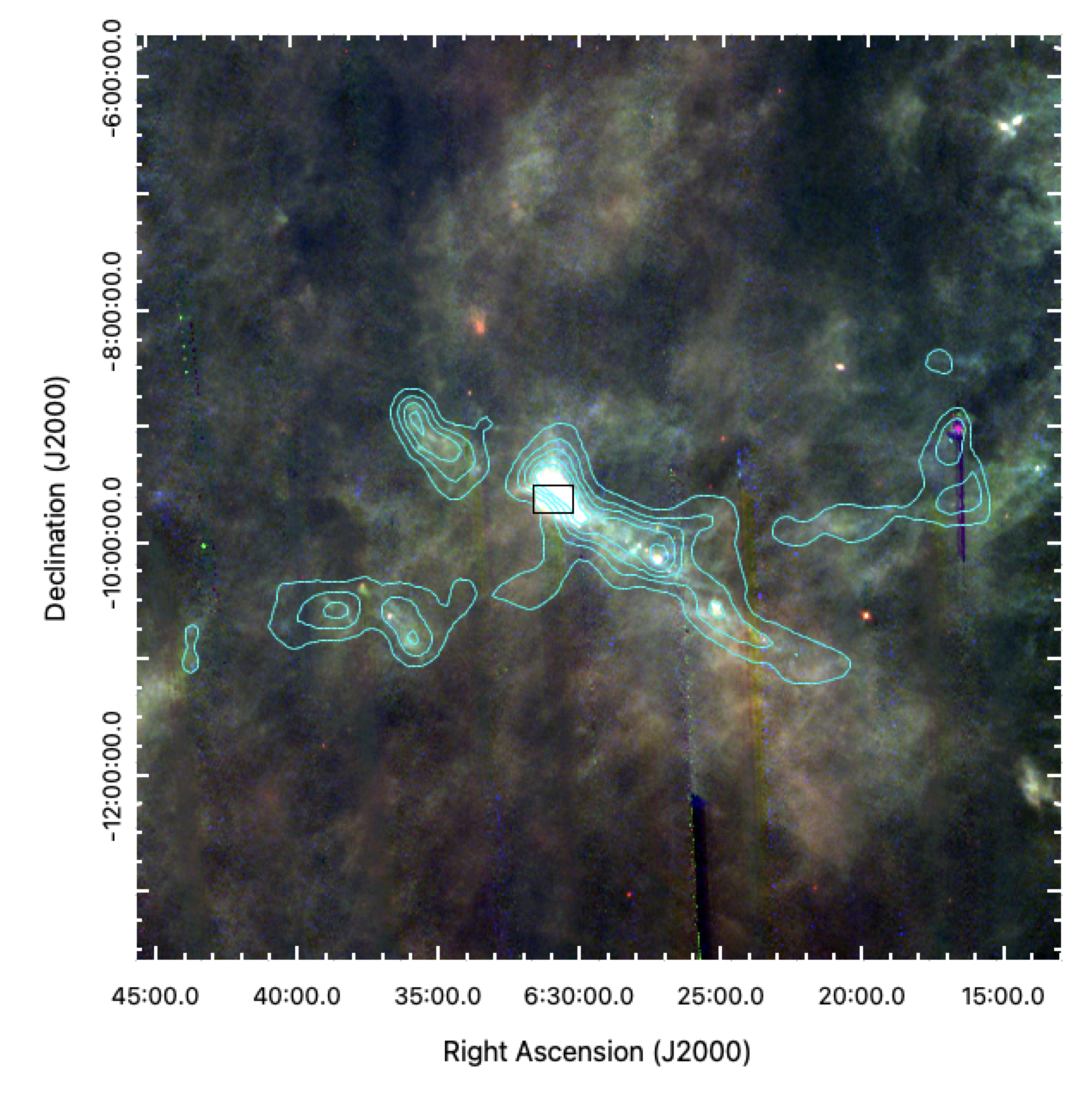}}
\caption{RGB composite image of the Crossbones filaments in far-infrared. Red is \textit{AKARI} N160 (160\,$\mu$m band), green is \textit{AKARI} WIDE-L (140\,$\mu$m band) and blue is \textit{AKARI} WIDE-S (90\,$\mu$m band). Contours in cyan are generated from the $^{12}$CO (J = 1--0) map provided by \citet{2024A&A...688A..54G}. The black square in the image represents the size of Figure~\ref{fig:Lagotis1}, Lagotis corresponds to the bright emission in far-infrared. Fragmented vertical lines present in the image are artefacts from \textit{AKARI} observations.}
\label{fig:Lone-Akari}
\end{figure*}

The \textit{AKARI} far-infrared All-Sky Survey \citep{2015PASJ...67...50D} comprises four far-infrared bands (65, 90, 140 and 160\,$\mu$m). \textit{AKARI} data show the gas and dust shroud surrounding the Lagotis \HII\ region and VdB-80 (Figure~\ref{fig:Lone-Akari}), and are used in conjunction with optical and radio images to create an RGB composite (Figure~\ref{RGB-Fig}). \textit{AKARI} survey data are useful for mapping the local area surrounding VdB-80, as well as the broader context of the Crossbones filaments. 

\section{Results and Discussion}
\label{sec:Results and Discussion}

\subsection{VdB-80 Radio-continuum Association}
\label{subsec:RCE}
\indent

Lagotis, the VdB-80-associated radio-continuum emission detected with \ac{EMU} is a roughly semi-circular feature with brighter emission in the western part (Figure~\ref{fig:Lagotis1}). 
It can be fitted by an ellipse
with size 2\farcm7$\times$2\farcm4, centred at RA(J2000)~6$^{\rm h}$30$^{\rm m}$53.5$^{\rm s}$ and Dec(J2000)~9$^\circ$39$^\prime$08\farcs6. 
The whole Lagotis structure has an integrated flux density of 
30.2$\pm$3.1\,mJy.\footnote{We measure the flux density and error following the same procedure as in~\citet{2022MNRAS.512..265F} and \citet{2023AJ....166..149F}.} 
The western side of the feature appears much brighter than the eastern, with integrated flux densities of 25.6$\pm$2.6\,mJy and 4.6$\pm$0.5\,mJy.

\begin{figure*}[ht]
\centerline{\includegraphics[trim=0 20 0 15 width=1.0\linewidth, keepaspectratio]{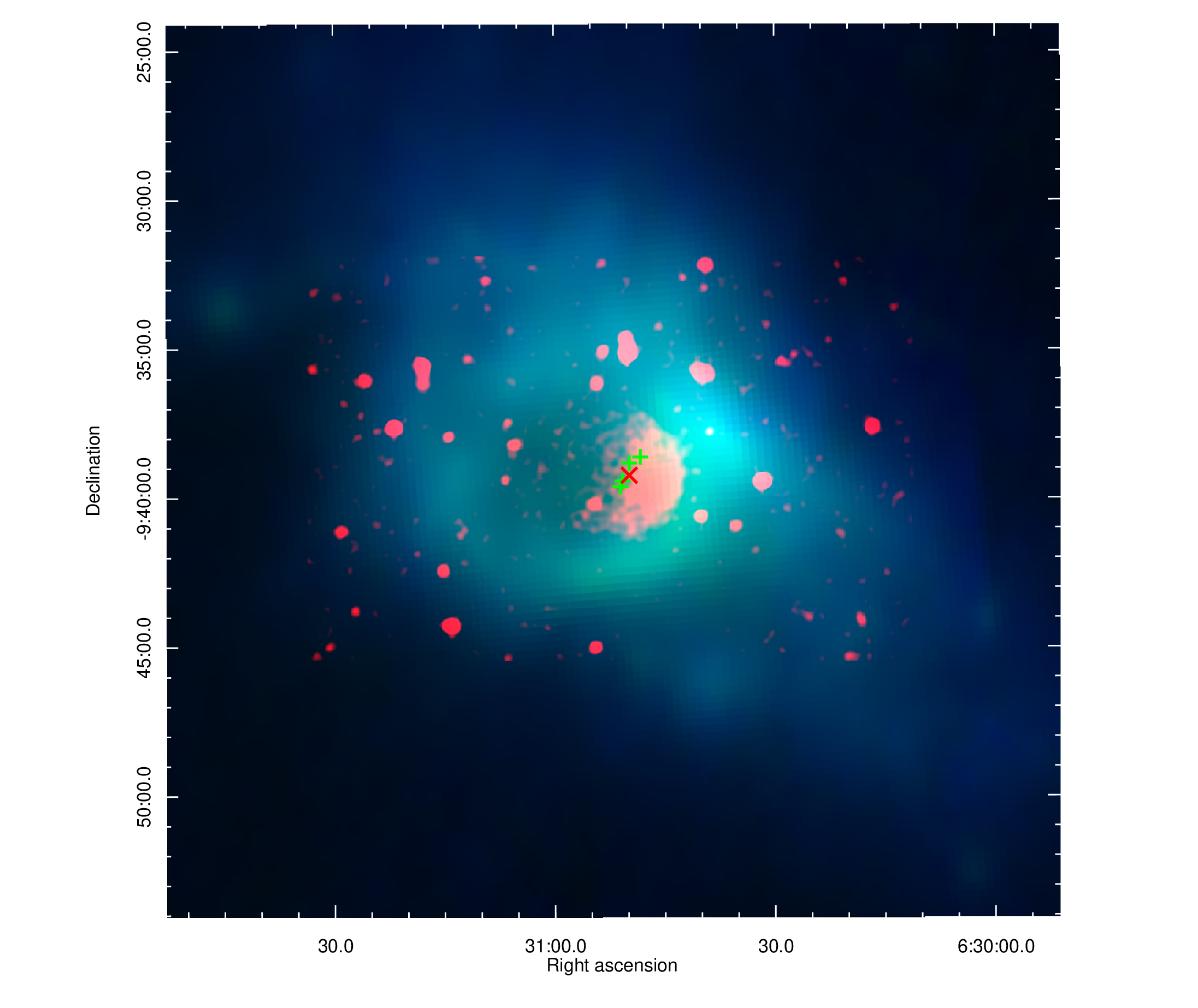}}
\caption{RGB image of Lagotis \HII\ region and VdB-80, where red is the \ac{EMU} radio image (943.5\,MHz), green is an \textit{AKARI} wide-S band (90 $\mu$m) image, and blue is an \textit{AKARI} wide-L band (140 $\mu$m) image. The red `X' denotes the star HD~46060, and the green crosses denote the other stars in the cluster (Table \ref{table:GaiaData}).}
\label{RGB-Fig}
\end{figure*}

Lagotis is detected in the \ac{NVSS} survey at 1.4\,GHz under the name NVSS 063048-094007 \citep{1998AJ....115.1693C}. Whilst this has been catalogued in \ac{NVSS}, there has been no study regarding this emission as an \HII\ region. The radio emission at 1.4\,GHz has a measured flux of 39.1$\pm$3.9 mJy, and so a spectral index can be estimated, following the spectral index definition $S\propto\nu^{\alpha}$~\citep{book1}. 
Convolving the \ac{EMU} data to the same resolution as \ac{NVSS} (beam size\,=\,45$\times$45\farcs, pixel size\,=\,10$\times$10\farcs) 
gives a spectral index of 0.7$\pm$0.5, a fairly flat spectrum which is indicative of thermal radio emission. 
This spectral index estimate allows us to estimate a surface brightness value of $\Sigma_{1\,\text{GHz}}$\,$\sim$\,7.3 $\times$10$^{-22}$ W m$^{-2}$ Hz$^{-1}$ sr$^{-1}$; assuming a distance of 960\,pc, this gives a luminosity at 1\,GHz of 
$\sim 3.6 \times 10^{12}$ W\,Hz$^{-1}$.
The approximately circular radio emission fits well inside the far-infrared emitting shell of gas and dust, as shown in Figure~\ref{RGB-Fig}.

\subsection{Distance and Size}
\label{subsec:Distance and Size}
\indent

\textit{Gaia} parallaxes of the young stars associated with the \ac{RN} VdB-80 and in the centre of the Lagotis \HII\ region (Table~\ref{table:GaiaData}) are taken from \textit{Gaia}~DR3 \citep{2016A&A...595A...1G, 2023A&A...674A...1G}. The derived distances range from 914 to 1127~pc. Since this range is much larger than the size of the stellar association \footnote{The young stars spread over about an arcminute on the sky, <1\,pc at $\sim 1$\,kpc.}, we treat all stars as lying at an error-weighted average distance of 960\,pc, with an uncertainty of about 100\,pc.

\citet{2005A&A...430..523W} estimated a distance to the Crossbones (and by extension Lagotis and VdB-80) of 465\,pc using Hipparcos parallaxes, significantly closer than the distance we derive using \textit{Gaia} data --- this disparity is considered further in section \ref{subsec:Crossbones}. We take the value of 960$\pm$100\,pc to be accurate as \textit{Gaia} parallax is more reliable than the Hipparcos parallax used for the closer distance derivation \citep{2021PASA...38....2A}.

Within the region of the radio-continuum emission, there are around 150 identified stars \citep{2016A&A...595A...1G, 2023A&A...674A...1G} with varying magnitudes and distances. The cluster of stars proposed to be associated with the VdB-80 \ac{RN} includes 12 of these, of which we use a subset of five stars, including HD~46060 (Table~\ref{table:GaiaData}); these appear to be central to the radio feature. These stars were chosen based on their central location in the radio--continuum feature, and within 40 arcseconds of the proposed host star; HD\,46060 (see section \ref{subsec:HD46060}.). The distance is taken to be the error-weighted average distance of the five stars, which puts VdB-80 at 960$\pm$100\,pc. From the radio angular size of 2\farcm7$\times$2\farcm4, we estimate the physical size to be 0.75$\times$0.67($\pm$0.09\,pc).

\ac{2MASS} extinction data show that the stellar cluster and VdB-80 are in the same plane and share a common distance. Table \ref{table:2MASSDat} shows near-IR \ac{2MASS} brightnesses for the 
same stars, along with colour excess ($J-H/H-K_s$) and visual extinction $A_V$, calculated using formulae from \citet{2006MNRAS.369.1901F} and \citet{2007MNRAS.378.1447F}. The extinction values show a few magnitudes' reddening of the stars, similar to those found towards the Crossbones \citep{2007MNRAS.378.1447F}, indicating that they are embedded in the cloud. There is a distinct correlation between 2MASS extinction and \textit{Gaia} distance 
(Tables~\ref{table:GaiaData} \& \ref{table:2MASSDat}): This suggests that HD~46060 is towards the front of the cloud (or may have cleared enough material to have lower extinction), while the other stars are more deeply embedded in the cloud behind HD~46060. The embedding of the stars underlines their association with the Crossbones cloud, and implies that the \textit{Gaia} distance can be used for the cloud. 

\begin{table*}
\caption{{Near-IR \ac{2MASS} $JHK_s$ magnitudes for the five star subset of the stellar cluster; and calculated near-IR colour excess ($J-H/H-K_s$) and visual extinction $A_V$, derived according to \citet{2006MNRAS.369.1901F}. The error in  extinction arises from the variation between $J-H$ and $H-K_s$ colours.}}
\vskip.25cm
\centerline{\begin{tabular}{lccccc}
 \hline
 Star Name & J & H & K$_s$ & J--H/H--K$_s$ & Extinction \\ 
 \hline\hline
 HD 46060 & 8.153 & 8.134 & 8.105 & 0.65 & 5.34$\pm$0.02 \\
 BD-09 1497 & 9.643 & 9.490 & 9.378 & 1.60 & 6.26$\pm$0.11 \\
 UCAC4 402-013961 & 10.808 & 10.478 & 10.317 & 2.00 & 6.96$\pm$0.20 \\
 UCAC4 402-013693 & 10.395 & 10.297 & 10.239 & 1.69 & 6.79$\pm$0.07 \\
 Gaia DR3 3002950320576187904 & 13.702 & 12.867 & 12.532 & 2.50 & 8.67$\pm$0.49 \\
 \hline
\end{tabular}}
\label{table:2MASSDat}
\end{table*}

\subsection{Crossbones}
\label{subsec:Crossbones}
\indent

The `Crossbones' in Monoceros is an `X'-shaped molecular cloud structure prominent in both $^{12}$CO and $^{13}$CO maps \citep{1986ApJ...303..375M,2004PASJ...56..313K}. It was originally identified as part of the Orion-Monoceros cloud complex by \citeauthor{1986ApJ...303..375M}: The Orion A and B clouds run approximately N--S, with northern and southern filaments running E--W. Crossbones appears to be part of the S filament, near where the Monoceros~R2 (Mon~R2) molecular cloud overlaps with the filament. At a distance of 800--900\,pc Mon~R2 lies far behind the Orion clouds (400--500\,pc), and is not associated. Crossbones has been taken to be a spatially coherent structure because the velocities of its parts appear coherent. \citeauthor{1986ApJ...303..375M},  noting that the stars associated with VdB-80 had estimated distances $\sim$800\,pc, suggested that the entire southern filament might be associated with Mon~R2, rather than Orion. \citet{2005A&A...430..523W} derived a distance of $\sim$460\,pc to the southern filament from \textit{Hipparcos} stellar parallaxes.

The Lagotis \HII\ region is embedded at the edge of the NE--SW arm of the Crossbones, close to the point where the arms cross (Figure~\ref{fig:Lone-Akari}). This can be seen in the far-IR images, in which a larger far-IR bright region envelops the \HII\ region, while the CO 1--0 emission has a cavity in the same place. The pressure of the \HII\ region and the UV flux of its star are excavating the cavity in the molecular cloud, and the dust around the cavity is being heated up to generate the far-IR emission (Figure~\ref{fig:Lone-Akari}). The calculated Lyman flux of HD\,46060 (see section \ref{subsec:HD46060}.)is consistent with the \HII\ region observed; this supports the argument that HD46060 and its neighbours not only illuminate VdB-80, but also power the Lagotis \HII\ region. Hence, the distance of these stars is the likely distance of this molecular gas structure. 

The most likely interpretation of the data is that Crossbones is not a contiguous structure, but a superposition of a NE--SW filamentary cloud and the southern filament of the Orion-Monoceros cloud complex. The population of YSOs found by \citet{2009ApJ...694.1423L} towards Crossbones are largely found towards the NE--SW arm, which also appears brighter in CO and far-IR. We therefore suggest that the NE--SW arm of Crossbones is a dense star-forming cloud at the distance of (and possibly associated with) Mon~R2; the SE--NW arm of Crossbones is simply part of the less dense southern Orion filament. The only evidence against this interpretation is the lack of velocity discontinuity between the two arms.


From the CO 1--0 data \citep{2024A&A...688A..54G}, we estimate an excitation temperature towards Lagotis of 12~K, compared to the 17~K found towards Mon~R2 \citep{2016MNRAS.461...22P}. 
Considering the beam dilution in the rather large \textit{Planck} beam, these temperature values are reasonably consistent.



\subsection{The Stellar Cluster and the \HII\ Region}
\label{subsec:HD46060}
\indent

The \textit{Gaia} proper motions of the main stars in the stellar cluster
(Table \ref{table:GaiaData}) suggest that the cluster is moving into the Crossbones filaments near the centre (Figure~\ref{fig:Lone-Akari}). Based on this movement, we give the \HII\ region the designation `Lagotis', as the stars are moving, or `burrowing' into the \HII\ region and the molecular cloud\footnote{Lagotis comes from the Latin \textit{Macrotis Lagotis}, the Australian Greater Bilby, \url{https://australian.museum/learn/animals/mammals/greater-bilby/}.}.

In the larger scope of the stellar cluster, there is a large population of stars at a similar distance ($\sim$1\,kpc); the majority of stars in this cluster have similar proper motion profiles to the Lagotis stars (Table~\ref{table:GaiaData}). Due to its central location and brightness, we propose that HD~46060 is the main driver of the \HII\ region. This star is the brightest in the central cluster and has generally been associated with the \ac{RN} itself. The distance from HD~46060 to the edge of the bright emission is 0.49~pc, and we take this to be the radius of the \HII\ region. We also assume the electron temperature of the \HII\ region to be 10$^{4}$\,K. Using the equations in \citet{1997pism.book.....D} and \citet{2016A&A...588A.143S}, we derive an electron density of 
26\,cm$^{-3}$, which can then be used to estimate an ionising flux of $10^{45.6}
\,$s$^{-1}$.

HD~46060 has been variously classified as a spectral type of B8 \citep{1966AJ.....71..990V, 1980BICDS..19...74O}, B3ne \citep{1968AJ.....73..233R}, \BII\ \citep{2000yCat.3214....0H, 2001KFNT...17..409K, 2012AstL...38..331A} and \BIII\ \citep{1972AJ.....77...17A}. Based on the most recent results, we adopt a spectral type of \BII. HD~46060 is about 4.5$\pm$1.5\,Myr old, \citep{2001A&A...377..845A}, so its ionising flux is expected to be higher than the \ac{ZAMS} value of $10^{44.7}$\,s$^{-1}$. \citet{1973AJ.....78..929P} gives a range of ionising fluxes for \BI\ that is consistent with our estimated ionising flux of $10^{45.6}\,$s$^{-1}$.

We therefore put forward a self-consistent interpretation in which HD~46060 is a \BII\ star with sufficient ionising flux to power the Lagotis \HII\ region with radius 0.49\,pc and electron density 26\,cm$^{-3}$. This gives rise to thermal radio continuum emission, consistent with the estimated flux density.



Because VdB-80 sits on the edge of the cloud, there may be a champagne flow \citep{1997A&A...326.1195C, 2014A&A...563A..39I} from the Lagotis \HII\ region that is not seen in the \ac{EMU} image, with supersonic movement of ionised gas caused by a steep pressure and density gradient at the edge of the cloud \citep{1979ApJ...233...85B}. This is consistent with the roughly hemispherical shape of the \HII\ region, and may be the cause for the dim eastern side of the radio--continuum emission. More sensitive observations at additional frequencies may be able to test this possibility.

\section{Conclusion}
\label{sec:Conclusion}
\indent

Sensitive \ac{EMU} observations have revealed radio-continuum emission at 943.5\,MHz toward a known \ac{RN}, VdB-80 \citep{1966AJ.....71..990V}. This is a unique feature for an object mainly known for its optical properties. The emission --- nicknamed \textit{Lagotis} --- is measured to be 30.2$\pm$0.3\,mJy, and is comprised of bright (Western side) and dim (Eastern side) portions with flux densities of 25.6$\pm$2.6\,mJy and 4.6$\pm$0.5\,mJy. 


The radio-continuum detection is a circular feature with the brighter side pointed inward toward the Crossbones cloud filaments, accompanied by a bright shroud of far-infrared emission and an extended shell of mid-infrared emission (Figures~\ref{fig:WISE34} \& \ref{RGB-Fig}). This indicates heating of the cloud surrounding the \ac{RN}, which is likely to be a byproduct of a \HII\ region heating up the gas and dust surrounding its shell. This shows that the features associated with the \ac{RN} and \HII\ region extend much further than shown in visible wavelengths.

VdB-80 and Lagotis are situated at the edge of the Crossbones molecular cloud structure. We find the \ac{RN} to be at a distance of 960$\pm$100\,pc based on \textit{Gaia} parallax values for a subset of stars in the stellar cluster association. From this distance, we estimate the size of the radio emission to be 0.75$\times$0.67($\pm$0.09\,pc), and we find that the Crossbones feature is most likely a superposition of two filamentary clouds associated with Orion (in the foreground) and Mon~R2 (in the background).

The proper motion (Table \ref{table:GaiaData}) of the stellar cluster associated with the \ac{RN} shows that the cluster is moving toward, and `into' the cloud. 
The proposed driver of the \HII\ region is HD~46060, a \BII\ type star at the center of the stellar cluster, with enough ionising output to power the \HII\ region, with  an estimated electron density of 26\,cm$^{-3}$. Irregularities in the radio feature as well as the far-infrared emitting shell may be generated from the other stars in the cluster with unknown spectral types. 

Objects with low surface brightness, such as Lagotis, are becoming more prominent in high sensitivity surveys such as \ac{EMU}. This discovery of the radio-continuum emission associated with an \ac{RN} and its analysis has indicated that HD~46060 both powers Lagotis and illuminates VdB-80. 
\ac{ASKAP} and other new-generation telescopes enable future observations and discoveries of \acp{RN} with radio-continuum emission.


\begin{acknowledgement}

This scientific work uses data obtained from Inyarrimanha Ilgari Bundara / the Murchison Radio-astronomy Observatory. We acknowledge the Wajarri Yamaji People as the Traditional Owners and native title holders of the Observatory site. CSIRO’s ASKAP radio telescope is part of the Australia Telescope National Facility (\url{https://ror.org/05qajvd42}). Operation of ASKAP is funded by the Australian Government with support from the National Collaborative Research Infrastructure Strategy. ASKAP uses the resources of the Pawsey Supercomputing Research Centre. Establishment of ASKAP, Inyarrimanha Ilgari Bundara, the CSIRO Murchison Radio-astronomy Observatory and the Pawsey Supercomputing Research Centre are initiatives of the Australian Government, with support from the Government of Western Australia and the Science and Industry Endowment Fund.

This work has made use of data from the European Space Agency (ESA) mission {\it Gaia} (\url{https://www.cosmos.esa.int/gaia}), processed by the {\it Gaia} Data Processing and Analysis Consortium (DPAC, \url{https://www.cosmos.esa.int/web/gaia/dpac/consortium}). Funding for the DPAC has been provided by national institutions, in particular the institutions participating in the {\it Gaia} Multilateral Agreement.
This publication makes use of data products from the Two Micron All Sky Survey, which is a joint project of the University of Massachusetts and the Infrared Processing and Analysis Center/California Institute of Technology, funded by the National Aeronautics and Space Administration and the National Science Foundation.
This research is based on observations with \textit{AKARI}, a JAXA project with the participation of ESA.

\end{acknowledgement}

\paragraph{Funding Statement}

MDF, GR and SL acknowledge Australian Research Council (ARC) funding through grant DP200100784. DU acknowledges the financial support provided by the Ministry of Science, Technological Development and Innovation of the Republic of Serbia through the contract 451-03-66/2024-03/200104 and for support through the joint project of the Serbian Academy of Sciences and Arts and Bulgarian Academy of Sciences  ``Optical search for Galactic and extragalactic supernova remnants''.

\paragraph{Competing Interests}

None.

\paragraph{Data Availability Statement}

\ac{EMU} data can be accessed through the \ac{CASDA} portal: \url{https://research.csiro.au/casda}. \ac{NVSS} data can be obtained from the \ac{NVSS} Postage Stamp Server: \url{https://www.cv.nrao.edu/nvss/postage.shtml}. All \textit{Gaia} \ac{DR3} data are obtained from the \textit{Gaia} Archive website: \url{https://gea.esac.esa.int/archive/}. Data from the \ac{2MASS} Point Source Catalogue are available from the NASA/IPAC Infrared Science Archive (IRSA): \url{https://irsa.ipac.caltech.edu/Missions/2mass.html}. \ac{WISE} data is available from the NASA/IPAC Infrared Science Archive (IRSA): \url{https://irsa.ipac.caltech.edu/Missions/wise.html}. Data from the \textit{AKARI} all-sky survey are available at the NASA/IPAC Infrared Science Archive (IRSA): \url{https://irsa.ipac.caltech.edu/data/AKARI/}.

\printendnotes

\bibliography{Lagotis}

\end{document}